\documentclass[%
 aip,
 jmp,%
 amsmath,amssymb,
 reprint,%
]{revtex4-1}

\usepackage{graphicx}
\usepackage{dcolumn}
\usepackage{bm}

\usepackage{epstopdf}

\begin{document}

\preprint{AIP/123-QED}

\title[Flow--turbulence interaction in reconnection]{Flow--turbulence interaction in magnetic reconnection}

\author{N. Yokoi}
\altaffiliation[Guest Researcher at the ]{National Astronomical Observatory of Japan (NAOJ)}
\altaffiliation[Guest Researcher at the ]{Nordic Institute for Theoretical Physics (NORDITA)}
\email{nobyokoi@iis.u-tokyo.ac.jp}
\affiliation{ 
Institute of Industrial Science, University of Tokyo, 4-6-1, Komaba, Meguro, Tokyo 153-8505, Japan
}%

\author{M. Hoshino}
\affiliation{%
Department of Earth and Planetary Science, Graduate School of Science, University of Tokyo, 7-3-1, Hongo, Bunkyo, Tokyo 113-0033, Japan}%

\date{\today}

\begin{abstract}
Roles of turbulence in the context of magnetic reconnection are investigated with special emphasis on the mutual interaction between flow (large-scale inhomogeneous structure) and turbulence. In order to evaluate the effective transport due to turbulence, in addition to the {\it intensity} information of turbulence represented by the turbulent energy, the {\it structure} information represented by pseudoscalar statistical quantities (helicities) is important. On the basis of the evolution equation, mechanisms that provide turbulence with cross helicity are presented. Magnetic-flux freezing in highly turbulent media is considered with special emphasis on the spatial distribution of the turbulent cross helicity. The cross-helicity effects in the context of magnetic reconnection are also investigated. It is shown that the large-scale flow and magnetic-field configurations favorable for the cross-helicity generation is compatible with the fast reconnection. In this sense, turbulence and large-scale structures promote magnetic reconnection mediated by the turbulent cross helicity.
\end{abstract}

\maketitle

\section{\label{sec:I}Introduction}

	Magnetic reconnection is one of the most important processes in plasma physics. It appears in a variety of phenomena in geo/astrophysics and fusion physics, which include dynamo, solar and stellar flares, magnetospheric substorms, disruptions in fusion devices, etc. In the framework of magnetohydrodynamics (MHD), the diffusion of magnetic field in magnetic reconnection can be described with a basic model proposed by Sweet\cite{swe1958} and Parker,\cite{par1957} where conservations of the magnetic flux, mass, and momentum are considered. However, Sweet--Parker mechanism of reconnection is too slow to explain typical time scales of eruptive processes observed in laboratory and in space such as solar flares. In order to alleviate this discrepancy due to the slowness of reconnection, several theoretical attempts for the fast reconnection, including the Petschek's model,\cite{pet1964} have been proposed. Turbulence is one of the candidate mechanisms that enhance the reconnection rate very much.

	Turbulence in magnetic reconnection has been investigated by several researchers. Analysis of magnetic-field observations in the Earth's magnetotail suggested the reconnection process in the current sheet has properties of turbulence excited by tearing vortices.\cite{hos1994} The effects of turbulence on magnetic reconnection were first investigated by Matthaeus and Lamkin using numerical simulations of pinch sheet. They showed that the electric fluctuations combined with the flow and current filament enhance the dissipation in the reconnection zone, and leads to the rapid reconnection.\cite{mat1985,mat1986} In order to dissolve the discrepancy between the typical length scale of phenomena and the dimension required for the current sheet, a fractal nature of the current sheet was suggested.\cite{taj1997} Lazarian and Vishniac examined the effects of stochastic field in reconnection and concluded that the presence of the stochastic components of the magnetic field enhances the reconnection rate drastically. They stressed that turbulence alters the relevant scales for reconnection almost independently of the microscopic dissipation mechanisms.\cite{laz1999}

	One of the primary effects of turbulence is enhancing the mixing or transport of the system, which is represented by the notion of turbulent transport such as the eddy viscosity, eddy diffusivity, anomalous resistivity, etc. For example, in the mean momentum equation in a turbulent flow, the eddy viscosity $\nu_{\rm{T}}$ enhances the effect of molecular viscosity $\nu$ as $\nu \to \nu + \nu_{\rm{T}}$ with spatiotemporal variation of $\nu_{\rm{T}}$. The enhanced transport due to turbulence leads to destruction of structure or reduction of spatiotemporal inhomogeneity. In the real-world phenomena, however, roles of turbulence are not so obvious or straightforward. Under some circumstances, a large-scale inhomogeneous structure is formed or sustained by turbulence itself. Representative situations are seen in turbulent dynamos, where large-scale magnetic fields are generated by some statistical properties of turbulent motion.\cite{mof1978,par1979,kra1981} This can be regarded as an example of general situations where breakage of symmetry leads to the formation or sustainment of large-scale inhomogeneous structure in turbulence. In the presence of symmetry breakage in turbulence, we may expect some effects that balance or counteract the primary turbulence effects. Here, enhanced transports due to turbulence are suppressed, and large-scale inhomogeneous structures are sustained or formed. Important question is how to formulate such breakage of symmetry in turbulence which leads to the structure formation.

	Most straightforward method for describing turbulent motions is to directly solve governing equations. However, for realistic turbulence of scientific interests with a huge Reynolds number, it is impossible in the foreseeable future to numerically solve all scales of motions ranging from the largest scales of system size to the smallest scales where the energy is dissipated into heat. In such a situation, modeling of turbulent motions provides a useful tool for investigating realistic turbulence. In practice of studying turbulence phenomena that appear in the geo/astrophysical and fusion sciences, one of the important tasks is how to choose quantities that appropriately describe the statistical properties of turbulence.
	
	In the studies of turbulence modeling, among several statistical quantities, the turbulent energy density (per unit mass) $k (\equiv \langle {{\bf{u}}'{}^2} \rangle /2)$ and its dissipation rate $\epsilon [\equiv \nu \langle {(\partial u'{}^a / \partial x^b)^2} \rangle]$ have often been adopted as descriptors of turbulence properties (${\bf{u}}'$: velocity fluctuation, $\nu$: kinematic viscosity).\cite{lau1972} The turbulent energy $k$ expresses how much fluctuation a turbulent flow has, and the energy dissipation rate $\epsilon$ expresses how much energy is transferred from large to small scales. In this sense, both $k$ and $\epsilon$ represent the {\it intensity} property of turbulence. The eddy-viscosity representation is supplemented by a model of $\nu_{\rm{T}}$ whose generic form is expressed by {\it intensity} measures of turbulence as $\nu_{\rm{T}} = f\{ {k,\epsilon} \}$. The most popular expression is $\nu_{\rm{T}} = C_\nu {k^2}/{\epsilon}$ ($C_\nu$: model constant, $k/\epsilon$: time scale of turbulence). However, it is well known in the engineering field that such a simple eddy-viscosity model completely fails if applied to some turbulent flows. The turbulent swirling flow is the representative case. A simple eddy-viscosity model is too dissipative to sustain the centerline dent profile of axial mean velocity that is experimentally observed in the turbulent swirling flow.\cite{kit1990,ste1995} In order to alleviate this discrepancy, anomalously small value of $C_\nu$ has often been adopted in the numerical simulations.\cite{kob1987} This fact implies that the description based on $k$ and $\epsilon$ is not enough and we have to take into account another turbulent statistical quantity that properly describes the {\it structure} property or breakage of mirrorsymmetry in the turbulent swirling pipe flow.\cite{yok1993}
	
	Large-scale rotation and magnetic field are often essential ingredients in the geo/astrophysical and fusion plasma phenomena. If we have large-scale rotational motion or magnetic field, symmetry of turbulence between the directions parallel and antiparallel to the rotation axis or magnetic field is broken. Such breakage of mirrorsymmetry is expected to be measured by pseudoscalar turbulent quantities, which change their sign under the inversion of the coordinate system. Since pseudoscalar vanishes in a mirrorsymmetric system, a finite value of pseudoscalar indicates the breakage of mirrorsymmetry.
	
	The turbulent cross helicity is a pseudoscalar which represents breakage of mirrorsymmetry, more precisely, asymmetry between the directions parallel and antiparallel to the large-scale magnetic field. In the presence of breakage of mirrorsymmetry in turbulence, the primary transports due to turbulence may be suppressed by the cross-helicity effects, and large-scale structures are formed and sustained even in the presence of a strong turbulence. These two effects: (i) generation of large-scale structures; and (ii) suppression of turbulent transport; are two sides of the same coin, the cross-helicity effect. Unlike the kinetic and magnetic helicities, the cross helicity can exist even in two dimensional configuration of magnetohydrodynamic plasmas. For physical interpretation of the cross-helicity effect and its application to the dynamo problems, the reader is referred to Refs.\ \onlinecite{yos1990,yok1999,yok2011b}
	
	Behavior of turbulent statistical quantities relevant to the transport coefficients provides essential information on the evolution of the large-scale structures. At the same time, how turbulent statistical quantities are spatially distributed and temporally evolve is determined by large-scale flow and magnetic-field structures or mean-field inhomogeneities. As is well known, the mean-velocity shear coupled with the Reynolds stress,$- \langle {u'{}^a u'{}^b} \rangle \partial U^a / \partial x^b$, is the main cause for the turbulence production in a hydrodynamic (HD) flow. In the MHD turbulent flow, as will be shown in Sec.~\ref{sec:II}, we have several mean-field inhomogeneities other than the mean velocity shear, which may contribute to the energy and cross-helicity generations.
	
\begin{figure}[htb]
\includegraphics[width=.35\textwidth]{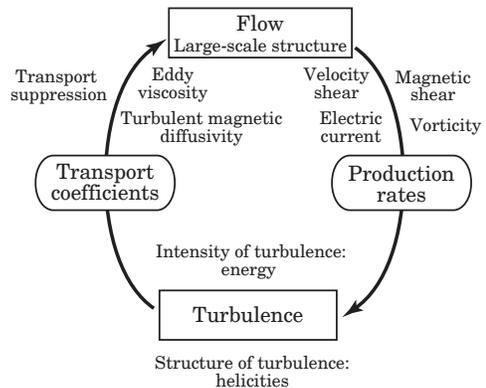}
\caption{\label{fig:flow_turbulence} Flow--turbulence relationship. Mean-field structures are determined by turbulence through the turbulent transport coefficients. Turbulence properties are determined by mean-field structures through the production rates of turbulent statistical quantities.}
\end{figure}

	Large-scale structures, which can be denoted in the broad sense as {\it flow} including the mean flow and global magnetic field, etc., are determined by {\it turbulence} through the turbulent transport coefficients. On the other hand, properties of turbulence are determined by the large-scale flow structures such as the mean velocity shear, vorticity, magnetic-field shear, electric current, etc. This mutual relationship between the flow and turbulence is schematically depicted in Fig.~\ref{fig:flow_turbulence}.

	If we see the magnetic reconnection phenomena from this perspective, important points are as follows. Magnetic reconnection is considered to be promoted by the enhanced magnetic diffusivity due to turbulence. If the spatiotemporal distributions of the turbulent statistical quantities (turbulent energy, cross helicity, etc.) relevant to the turbulent transports are favorable for the magnetic reconnection, the rate of reconnection would be increased. The spatiotemporal distributions of these turbulent quantities obey the balance of the production, dissipation, and transport rates of these statistical quantities. The production mechanism, the most important ingredient, is determined by the large-scale flow structures such as the inflow profiles, global magnetic-field configurations, etc. If the large-scale structures sustained and formed during the magnetic reconnection process are consistent with the favorable distributions of the transport coefficients for the magnetic reconnection, the reconnection would be promoted. Otherwise, the reconnection rate should be reduced. Therefore, it is the production mechanism of the turbulent statistical quantities due to large-scale structures that should be examined intensively. In this paper, we focus our attention on the turbulent cross-helicity effects, and address the magnetic reconnection from the viewpoint of the mutual relationship between the flow structure and turbulence mediated by the cross helicity.

	This paper is organized as follows. In Sec.~\ref{sec:II}, the evolution equation of the turbulent cross helicity is presented. On the basis of this equation, main mechanisms that provide turbulence with the cross helicity are examined. In Sec.~\ref{sec:III}, the cross-helicity effects in the mean magnetic induction equation are treated with special emphasis on the conditions for the magnetic-flux freezing and the magnetic reconnections. In Sec.~\ref{sec:IV}, the cross-helicity effects in the mean momentum equation are considered. The mean flow configuration affects the reconnection rates very much. By considering the flow induced by the cross-helicity effects, we discuss the influence of the cross helicity in turbulence on the magnetic reconnections. Summary and concluding remarks are presented in Sec.~\ref{sec:V}.

\section{\label{sec:II}Cross-helicity evolution}

\subsection{\label{sec:II-A}Reynolds stress and turbulent electromotive force}

	Effects of turbulence on the mean fields in magnetohydrodynamics are represented by the Reynolds (and turbulent Maxwell) stress $\mbox{\boldmath${\cal{R}}$} = \langle {{\bf{u}}'{\bf{u}}' - {\bf{b}}'{\bf{b}}'} \rangle$ in the mean velocity (${\bf{U}}$) equation and by the turbulent electromotive force ${\bf{E}}_{\rm{M}} = \langle {{\bf{u}}' \times {\bf{b}}'} \rangle$ in the mean magnetic-field (${\bf{B}}$) equation (${\bf{b}}'$: magnetic-field fluctuation). Note that, in this paper, we adopt the Alfv\'{e}n speed unit for the magnetic field etc.\ (${\bf{b}} = {\bf{b}}_\ast / \sqrt{\mu \rho}$, ${\bf{b}}_\ast$: magnetic field measured in the original physical unit, $\mu$: magnetic permeability, $\rho$: density). In the case of compressible flows, we have several turbulent transport mechanisms that are intrinsically connected to the density variation. For example, density variation along the mean magnetic field gives an important contribution to the turbulent cross-helicity generation. Effects of compressiblity or variable density can be incorporated into the cross-helicity arguments in the context of turbulent dynamos and turbulent transport suppression. However, such arguments will make the description much more complicated. At low plasma beta, incompressibility is considered to be a good approximation although the plasma beta becomes about one at the current-sheet region. Considering that the essential physics of magnetic reconnection is not so different in incompressible and compressible cases, here in this paper, as most work following the original Sweet--Parker model, we confine our arguments to the incompressible framework. This treatment never denies the possible importance of compressibility in the magnetic reconnection study especially in the astrophysical context. For some examples of the theoretical extensions to the compressible or variable-density case, the reader is referred to Refs.~\onlinecite{yos2003} and \onlinecite{yok2007}.
	
	As we see in Appendix A, from a statistical analytical theory for inhomogeneous turbulence applied to magnetohydrodynamic (MHD) turbulence, $\mbox{\boldmath${\cal{R}}$}$ and ${\bf{E}}_{\rm{M}}$ are expressed as\cite{yos1990}
\begin{subequations}\label{eq:Re_str}
\begin{eqnarray}
	{\cal{R}}^{\alpha\beta} 
	&\equiv& \left\langle {
		u'{}^\alpha u'^\beta - b'^\alpha b'^\beta
	} \right\rangle
	\label{eq:Re_str_def}\\
	& = & \frac{2}{3} K_{\rm{R}} \delta^{\alpha\beta}
	- \nu_{\rm{K}} {\cal{S}}^{\alpha\beta}
	+ \nu_{\rm{M}} {\cal{M}}^{\alpha\beta},
	\label{eq:Re_str_exp}
\end{eqnarray}
\end{subequations}
\begin{subequations}\label{eq:E_M}
\begin{eqnarray}
	{\bf{E}}_{\rm{M}}
	&\equiv& \left\langle {{\bf{u}}' \times {\bf{b}}'} \right\rangle
	\label{eq:E_M_def}\\
	&=&  - \beta {\bf{J}} + \alpha {\bf{B}} + \gamma \mbox{\boldmath$\Omega$},
	\label{eq:E_M_exp}
\end{eqnarray}
\end{subequations}
where $K_{\rm{R}} (\equiv \langle {{\bf{u}}'{}^2 - {\bf{b}}'{}^2} \rangle/2)$ is the turbulent MHD residual energy, $\mbox{\boldmath$\cal{S}$}$ and $\mbox{\boldmath$\cal{M}$}$ are the  strain rates of the mean velocity and magnetic field, respectively, ${\bf{J}}$ the mean electric-current density, $\mbox{\boldmath$\Omega$} (= \nabla \times {\bf{U}})$ the mean vorticity. Here, $\nu_{\rm{K}}$, $\nu_{\rm{M}}$, $\alpha$, $\beta$ and $\gamma$ are the transport coefficients whose spatial distributions are determined by the properties of turbulence. In Eqs.~(\ref{eq:Re_str_exp}) and (\ref{eq:E_M_exp}), the eddy viscosity and the turbulent magnetic diffusivity, $\nu_{\rm{K}}$ and $\beta$, are expressed in terms of the turbulent energy $K (\equiv \langle {{\bf{u}}'{}^2 + {\bf{b}}'{}^2} \rangle/2)$. The $\mbox{\boldmath$\cal{M}$}$-related transport coefficient  $\nu_{\rm{M}}$ in Eq.~(\ref{eq:Re_str_exp}) and the $\mbox{\boldmath$\Omega$}$-related coefficient $\gamma$ in Eq.~(\ref{eq:E_M_exp}) are expressed in terms of the turbulent cross helicity (velocity--magnetic-field correlation in turbulence) $W (\equiv \langle {{\bf{u}}' \cdot {\bf{b}}'} \rangle)$. Then, it is important to examine the evolution equations of the turbulent energy and cross helicity, $K$ and $W$. In the context of turbulent transport suppression and structure formation, in particular, the production mechanisms of $W$ should be intensively investigated.
 
	Turbulent transports such as the eddy viscosity, turbulent resistivity, dynamo coefficients, etc.\ are determined by statistical properties of turbulence.   A simplest way to treat such statistical properties is adopting one-point turbulent statistical quantities and consider the evolution of these statistical quantities. In magnetohydrodynamic (MHD) turbulence, the turbulent MHD energy $K$ and the turbulent cross helicity $W$ defined by
\begin{equation}
	K \equiv \frac{1}{2} \left\langle {{\bf{u}}'{}^2 + {\bf{b}}'{}^2} \right\rangle,
	\label{eq:K_def}
\end{equation}
\begin{equation}
	W \equiv \left\langle {{\bf{u}}' \cdot {\bf{b}}'} \right\rangle
	\label{eq:W_eq}
\end{equation}
are most important ones among several statistical quantities since they are directly connected to the transport coefficients appearing in $\mbox{\boldmath$\cal{R}$}$, $\nu_{\rm{K}}$ and $\nu_{\rm{M}}$, and the ones in ${\bf{E}_{\rm{M}}}$, $\beta$ and $\gamma$. For detailed description of the transport equations of several statistical quantities, the reader is referred to Refs.~\onlinecite{yok2008,yok2011}.

\subsection{\label{sec:II-B}Equations of turbulent MHD energy and cross helicity}

	In order to see how and how much energy and cross helicity exist in turbulence, we have to consider the transport equations for $K$ and $W$. From equations of fluctuation velocity and magnetic fields, ${\bf{u}}'$ and ${\bf{b}}'$, equations for the turbulent MHD energy (density) and turbulent cross helicity (density) are written as
\begin{equation}
	\frac{DG}{Dt}
	\equiv \left( {\frac{\partial}{\partial t} + {\bf{U}}\cdot \nabla} \right) G
	= P_G - \varepsilon_G + T_G
	\label{eq:G_eq}
\end{equation}
with $G = (K, W)$. Here $P_G$, $\varepsilon_G$, and $T_G$ are the production, dissipation, and transport rates of $G$. They are defined by
\begin{subequations}\label{eq:K_evol}
\begin{equation}
	P_K
	= - {\cal{R}}^{ab} \frac{\partial U^b}{\partial x^a}
	- {\bf{E}}_{\rm{M}} \cdot {\bf{J}},
	\label{eq:P_K}
\end{equation}
\begin{equation}
	\varepsilon_K
	= \nu \left\langle {
		\frac{\partial u'{}^b}{\partial x^a}
		\frac{\partial u'{}^b}{\partial x^a}
	} \right\rangle
	+ \eta \left\langle {
		\frac{\partial b'{}^b}{\partial x^a}
		\frac{\partial b'{}^b}{\partial x^a}
	} \right\rangle
	\equiv \varepsilon,
	\label{eq:eps_K}
\end{equation}
\begin{equation}
	T_K
	= {\bf{B}} \cdot \nabla W
	+ \nabla \cdot {\bf{T}}'_K,
	\label{eq:T_K}
\end{equation}
\end{subequations}
\begin{subequations}\label{eq:W_evol}
\begin{equation}
	P_W
	= - {\cal{R}}^{ab} \frac{\partial B^b}{\partial x^a}
	- {\bf{E}}_{\rm{M}} \cdot \mbox{\boldmath$\Omega$},
	\label{eq:P_W}
\end{equation}
\begin{equation}
	\varepsilon_W
	= \left( {\nu + \eta} \right) 
	\left\langle {
		\frac{\partial u'{}^b}{\partial x^a}
		\frac{\partial b'{}^b}{\partial x^a}
	} \right\rangle,
	\label{eq:eps_W}
\end{equation}
\begin{equation}
	T_W
	= {\bf{B}} \cdot \nabla K
	+ \nabla \cdot {\bf{T}}'_W
	\label{eq:T_W}
\end{equation}
\end{subequations}
($\eta$: magnetic diffusivity). In Eqs.~(\ref{eq:P_K}) and (\ref{eq:P_W}), $\mbox{\boldmath${\cal{R}}$}$ and ${\bf{E}}_{\rm{M}}$ are the the Reynolds stress and the turbulent electromotive force defined by Eqs.~(\ref{eq:Re_str_def}) and (\ref{eq:E_M_def}), respectively. In Eqs. (\ref{eq:T_K}) and (\ref{eq:T_W}), ${\bf{T}}'_K$ and ${\bf{T}}'_W$ are the higher order correlations of fluctuations whose exact expressions are suppressed here.

\subsection{\label{sec:II-C}Cross-helicity generation}

	We see from Eq.~(\ref{eq:G_eq}) with Eq.~(\ref{eq:W_evol}) that there are three main mechanisms that provide turbulence with the cross helicity. They are the effects related to (i) the Reynold stress, $- {\cal{R}}^{ab} (\partial B^b/\partial x^a)$; (ii) the turbulent electromotive force, $- {\bf{E}}_{\rm{M}} \cdot \mbox{\boldmath$\Omega$}$; and (iii) inhomogeneity along the mean magnetic field, ${\bf{B}} \cdot \nabla K$. Among these three, the Reynolds-stress- and turbulent-electromotive-force-related productions are associated with the drain of the mean-field cross helicity ${\bf{U}} \cdot {\bf{B}}$. In this sense, they are related to the cascading property of the cross helicity. On the other hand, the turbulent cross-helicity generation due to the inhomogeneity along the mean magnetic field is not directly related to such cascade processes but arises from a kind of transport with asymmetry of the boundary conditions. In this sense, the inhomogeneity along the mean magnetic field may play a special role in the turbulent cross-helicity generation in the context of externally injecting cross helicity.
	
	In the following, we refer to some typical situations in which each of these generation mechanisms may work.

\subsubsection{\label{sec:II-C-1}Production due to Reynolds stress}

	The Reynolds stress coupled with the mean magnetic shear gives a cross-helicity generation. Substituting Eq.~(\ref{eq:Re_str_def}) into the first term of Eq.~(\ref{eq:P_W}), we have a production due to the Reynolds stress as
\begin{equation}
	- {\cal{R}}^{ab} \frac{\partial B^b}{\partial x^a}
	= \frac{1}{2} \nu_{\rm{K}} {\cal{S}}^{ab} {\cal{M}}^{ab}
	- \frac{1}{2} \nu_{\rm{M}} ({\cal{M}}^{ab})^2.
	\label{eq:Pw_Re}
\end{equation}
Since the sign of $\nu_{\rm{M}}$ is the same as that of $W$ as we see from Eqs.~(\ref{eq:nu_K_beta_rel}) and (\ref{eq:gamma_model}), the second or $\nu_{\rm{M}}$-related term always destroys the cross helicity regardless of the sign of $W$. So, as the production mechanism of the turbulent cross helicity, we write only the $\nu_{\rm{K}}$-related term as
\begin{equation}
	- {\cal{R}}^{ab} \frac{\partial B^b}{\partial x^a}
	= \frac{1}{2} \nu_{\rm{K}} {\cal{S}}^{ab} {\cal{M}}^{ab}.
	\label{eq:Pw_Re_SM}
\end{equation}

\begin{figure}[htb]
\includegraphics[width=.45\textwidth]{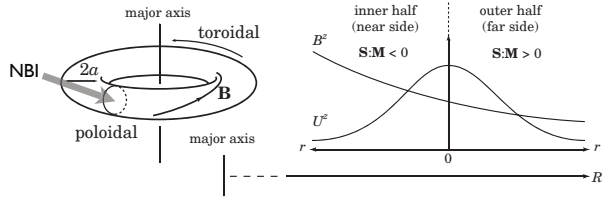}
\caption{\label{fig:pro_W_Re_str} Cross-helicity production due to the neutral beam injection (NBI). The Reynolds stress coupled with the mean magnetic-field shear leads to the  positive or negative cross-helicity generation.}
\end{figure}

	This shows that a positive cross helicity is generated if the mean velocity and magnetic shears have the same signs. A typical situation may be caused by a neutral beam injection (NBI) in the toroidal direction to tokamak devices in fusion plasmas. In toroidal confinement plasmas, the toroidal magnetic field decreases as the major radius of plasma increases ($B^\phi \propto 1/R$, $R$: distance from the major axis). If a neutral beam is injected in the center (minor axis) region, the toroidal velocity is maximum at the center of the minor radius ($r$: distance from the minor axis) and decreases as the minor radius of plasma increases. In this situation, we have negative velocity and magnetic shears ($\partial U^\phi /\partial r < 0, \partial B^\phi /\partial r < 0$) in the half region far from the major axis (far side), and negative velocity and positive magnetic shears ($\partial U^\phi /\partial r < 0, \partial B^\phi /\partial r > 0$) in the half region near the major axis (near side) as shown in Fig.~\ref{fig:pro_W_Re_str}. This suggests that we have positive and negative cross-helicity generations in the outer- and inner-half regions of plasma, respectively.

\subsubsection{\label{sec:II-C-2}Production due to turbulent electromotive force,}

\begin{figure}[htb]
\includegraphics[width=.45\textwidth]{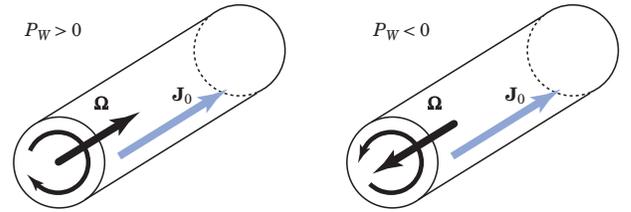}
\caption{\label{fig:pro_W_E_M} Cross-helicity production due to the coupling of the large-scale electric current and vorticity. If the large-scale vorticity associated with the poloidal rotation is parallel (or antiparallel) to the plasma current, a positive (or negative) cross helicity is generated.}
\end{figure}

	The turbulent electromotive force coupled with the mean vorticity $\mbox{\boldmath$\Omega$}$ gives a cross-helicity generation. Substituting Eq.~(\ref{eq:E_M_exp}) into the second term of Eq.~(\ref{eq:P_W}), we have a production due to the turbulent electromotive force as
\begin{eqnarray}
	- {\bf{E}}_{\rm{M}} \cdot \mbox{\boldmath$\Omega$}
	&=& \beta {\bf{J}} \cdot \mbox{\boldmath$\Omega$}
	- \gamma \mbox{\boldmath$\Omega$}^2
	- \alpha {\bf{B}} \cdot \mbox{\boldmath$\Omega$}.
	\label{eq:Pw_Em}
\end{eqnarray}
Since the sign of $\gamma$ is the same as that of $W$ as seen in Eq.~(\ref{eq:gamma_model}), the second or $\gamma$-related term in Eq.~(\ref{eq:Pw_Em}) works for the destruction of the turbulent cross helicity regardless of the sign of $W$. In tokamak plasmas, we have a very strong plasma current ${\bf{J}}_0$. In such a case, the turbulent cross-helicity generation due to the electromotive force may be expressed as
\begin{equation}
	- {\bf{E}}_{\rm{M}} \cdot \mbox{\boldmath$\Omega$} 
	\simeq \beta {\bf{J}}_0 \cdot \mbox{\boldmath$\Omega$}
	\label{eq:Pw_Em_JOmega}
\end{equation}
since $|{\bf{J}}_0| \gg |\mbox{\boldmath$\Omega$}|$. This shows that the positive or negative turbulent cross helicity is generated if the mean vorticity is aligned with the mean electric current:
\begin{equation}
	\left\{ {
	\begin{array}{lll}
	{\bf{J}}_0 \cdot \mbox{\boldmath$\Omega$} >0 &
	\rightarrow & P_W > 0,\\
	\rule{0.ex}{3.ex}
	{\bf{J}}_0 \cdot \mbox{\boldmath$\Omega$} <0 &
	\rightarrow & P_W < 0.
	\end{array}
	} \right.
	\label{eq:J_Omega_P_W}
\end{equation}
In tokamak plasmas, we have a strong plasma current in the toroidal direction, $J^z$ ($z$: toroidal direction in the cylindrical approximation for a torus geometry). In the presence of poloidal rotation $U^\theta$, we have a mean toroidal vorticity associated with this poloidal rotation, $\Omega^z [= (1/r)(\partial / \partial r) r U^\theta]$. Then, it is highly probable that we have a generation of positive or negative turbulent cross helicity depending on the direction of poloidal velocity (Fig.~\ref{fig:pro_W_E_M}). 

	As for the cause of poloidal rotation, we may consider several spontaneous plasma rotation associated with the improved confinement mode, such as local poloidal rotation near the plasma edge in H mode, global poloidal rotation near the internal transport barrier in the negative magnetic shear mode, etc. If we use a NBI with a tilted angle, we have a poloidal rotation in addition to the toroidal rotation. However, direct observation of the turbulent cross helicity in high temperature fusion devices is still very difficult. To validate this type of cross-helicity generation, a numerical simulation of cylindrical geometry mimicking the tokamak plasmas was performed.\cite{yok1999} In the work, it was shown that a poloidal plasma rotation, which corresponds to the mean vorticity in the toroidal direction, coupled with the plasma current contributes to the cross-helicity generation in turbulence. It was also shown that if the direction of poloidal rotation is changed, the sign of the generated cross helicity changes.

\subsubsection{\label{sec:II-C-3}Production due to inhomogeneity along the mean magnetic field}

\begin{figure}[htb]
\includegraphics[width=.45\textwidth]{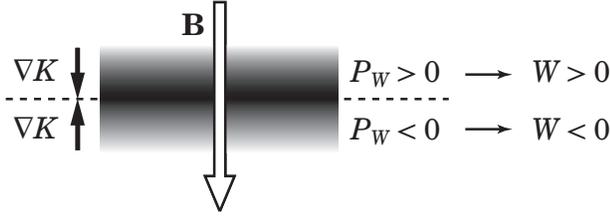}
\caption{\label{fig:pro_W_inhomo} Cross-helicity production due to the energy inhomogeneity along the large-scale magnetic field.}
\end{figure}

	If the turbulent MHD energy $K (\equiv \langle {{\bf{u}}'{}^2 + {\bf{b}}'{}^2} \rangle /2) $ is inhomogeneously distributed along the mean magnetic field ${\bf{B}}$, the cross helicity may be generated in turbulence. A positive (or negative) cross helicity is produced for $\nabla K$ parallel (or antiparallel) to ${\bf{B}}$:
\begin{equation}
	\left\{ {
	\begin{array}{lll}
		{\bf{B}} \cdot \nabla K > 0 & 
		\rightarrow &P_W > 0,\\
		\rule{0.ex}{3.ex}
		{\bf{B}} \cdot \nabla K < 0 & 
		\rightarrow &P_W < 0.
	\end{array}
	} \right.
	\label{eq:B_grad_K}
\end{equation}
It is ubiquitous in astrophysical phenomena that an ambient magnetic field threads through a plasma gas disk. If intensity of turbulence is not homogeneously distributed in the direction perpendicular to the disk, we have inhomogeneity of turbulence along the mean magnetic field. This mechanism may play a very important role in cross-helicity generation in astrophysical turbulence.\cite{yok1996} We consider a situation where a large-scale magnetic field threads through a disk from above to below, and the intensity of turbulence is highest at the midplane and decreases with the distance from the midplane (Fig.~\ref{fig:pro_W_inhomo}). In such a case, we have positive and negative cross-helicity generations in the upper- and lower-half regions of the disk. Consequently, we have positive and negative distributions of the cross helicity in the upper and lower halves of the disk as
\begin{equation}
	\left\{ {
	\begin{array}{ll}
		W > 0\;\;\; \mbox{for}\;\;\; z > 0,\\
		W < 0\;\;\; \mbox{for}\;\;\; z < 0.
	\end{array}
	} \right.
	\label{eq:spatio_dis_W}
\end{equation}

	We should stress that, in this situation, the intensity of turbulence is maximum at the midplane so that the gradient of turbulent energy, $\nabla K$, is zero and changes its sign there. As this result, there is no cross-helicity generation at the midplane, and the magnitude of the cross helicity should vanish there. This is consistent with a basic feature of pseudoscalars, representing the breakage of mirrorsymmetry in turbulence.

\section{\label{sec:III}Cross-helicity effects in magnetic induction}

\subsection{\label{sec:III-A}Magnetic-flux freezing in turbulence}

	The primary effect of turbulence in the magnetic induction equation is destruction of magnetic field, which is expressed by the turbulent magnetic diffusivity or anomalous resistivity $\beta$. If we have no field-generation effects such as the helicity or cross-helicity effect, the turbulent electromotive force is expressed as
\begin{equation}
	{\bf{E}}_{\rm{M}} 
	\equiv \left\langle {{\bf{u}}' \times {\bf{b}}'} \right\rangle
	= - \beta {\bf{J}}.
	\label{eq:E_M_only_beta}
\end{equation}	
Substituting this into the mean induction equation, we have
\begin{equation}
	\frac{\partial{\bf{B}}}{\partial t}
	= \nabla \times \left( {{\bf{U}} \times {\bf{B}}} \right)
	- \nabla \times \left[ {
		\left( {\eta + \beta} \right) \nabla \times {\bf{B}}
	} \right].
	\label{eq:mean_B_eq_only_beta}
\end{equation}	
This shows that the effective magnetic diffusivity is enhanced by turbulence as
\begin{equation}
	\eta \rightarrow \eta + \beta.
	\label{eq:enhance_resist}
\end{equation}
We should note that at large magnetic Reynolds number ($Rm \gg 1$), the turbulent magnetic diffusivity is in general much larger than the molecular counterpart as
\begin{equation}
	{\beta}/{\eta} \sim Rm \gg 1.
	\label{eq:large_Rm}
\end{equation}

	In the absence of any field-generation mechanisms, the mean induction equation is written as
\begin{equation}
	\frac{\partial{\bf{B}}}{\partial t}
	= \nabla \times \left( {{\bf{U}} \times {\bf{B}}} \right)
	- \nabla \times \left[ {\beta \nabla \times {\bf{B}}} \right].
	\label{eq:mean_B_eq_beta}
\end{equation}
In this case, the magnetic-flux freezing to the plasma motion does not occur since $\beta$ is so large that the effective magnetic diffusivity can not be neglected.

	On the other hand, if we have some field-generation mechanisms such as the helicity and the cross-helicity effects, the turbulent electromotive force is expressed as
\begin{equation}
	{\bf{E}}_{\rm{M}} 
	\equiv \left\langle {{\bf{u}}' \times {\bf{b}}'} \right\rangle
	= - \beta {\bf{J}}
	+ \alpha {\bf{B}}
	+ \gamma \mbox{\boldmath$\Omega$}.
	\label{eq:E_M_full}
\end{equation}
Then, the mean induction equation reads
\begin{eqnarray}
	\frac{\partial{\bf{B}}}{\partial t}
	&=& \nabla \times \left( {{\bf{U}} \times {\bf{B}}} \right)
	- \nabla \times \left[ {
		\left( {\eta + \beta} \right) \nabla \times {\bf{B}}
	} \right]\nonumber\\
	& & + \nabla \times \left( {
	\alpha {\bf{B}}
	+ \gamma \mbox{\boldmath$\Omega$}
	} \right).
	\label{eq:mean_B_eq_E_M_full}
\end{eqnarray}
The third term on the right-hand side of Eq.~(\ref{eq:mean_B_eq_E_M_full}) represents the field-generation mechanisms due to the turbulent residual helicity coupled with the mean magnetic field, $\alpha {\bf{B}}$, and the turbulent cross helicity coupled with the mean vorticity, $\gamma \mbox{\boldmath$\Omega$}$. If such field-generation mechanisms are large enough, they may balance the turbulent magnetic diffusivity effect, $\beta {\bf{J}}$, as
\begin{equation}
	\alpha {\bf{B}}
	+ \gamma \mbox{\boldmath$\Omega$}
	\sim \beta {\bf{J}}.
	\label{eq:balance_in_E_M}
\end{equation}
In this case, the contribution from the turbulent electromotive force is very small, and we have the mean induction equation as
\begin{equation}
	\frac{\partial{\bf{B}}}{\partial t}
	= \nabla \times \left( {{\bf{U}} \times {\bf{B}}} \right)
	+ \eta \nabla^2 {\bf{B}}.
	\label{eq:mean_B_eq_supp}
\end{equation}
This equation has exactly the same form as the induction equation for the instantaneous magnetic field ${\bf{b}}$. Then, for the flow with very high magnetic Reynolds number ($Rm \gg 1$), the mean magnetic field ${\bf{B}}$ is subject to the first term of the right-hand side of Eq.~(\ref{eq:mean_B_eq_supp}). Again, we have a magnetic-flux freezing to the plasma motion.

	Although this situation looks very similar to the magnetic-flux freezing in the laminar case, it is completely different from the laminar one. First, in Eq.~(\ref{eq:mean_B_eq_supp}), we have very strong turbulence or fluctuating components of the velocity and magnetic fields. This magnetic-flux freezing occurs only at very high Reynolds and magnetic Reynolds numbers flow. In this sense, we may call this the turbulent magnetic-flux freezing. Second point is that since this state is based on an equilibrium condition~(\ref{eq:balance_in_E_M}), the ``turbulent magnetic-flux freezing'' highly depends on the mean-field configurations and spatial and temporal distribution of turbulent quantities. If the balance relation between the field destruction and generation breaks, the picture of magnetic-flux freezing can not hold any more.
	
	If the field-generation mechanisms are absent, the field-destruction due to turbulence prevails. A typical situation may occur on the midplane of a plasma gas disk or equatorial plane in a spherical convection zone. Recalling the fact that pseudoscalar turbulent quantities are descriptors of breakage of mirrorsymmetry, it is quite natural these pseudoscalar quantities are distributed antisymmetricaly with respect to the midplane. For such spatial distributions, the magnitudes of pseudoscalar quantities vanish on the midplane, and very small in the vicinity of it. As a consequence of this disappearance of psedoscalars, a strong effective magnetic diffusivity may cause the magnetic reconnection in the close vicinity of the midplane.

\subsection{\label{sec:III-B}$\Omega$ and cross-helicity effects}

\subsubsection{\label{sec:III-B-1}$\Omega$ effect}

	In the study of dynamo, the so-called $\Omega$ effect has been considered as a basic ingredient of the dynamo process. The $\Omega$ effect contains several different physical processes in it. In the context of the toroidal-field generation from the poloidal magnetic field, at least three processes: (i) magnetic-flux freezing; (ii) appropriate differential rotation; (iii) magnetic-field reconnection; should be included in the $\Omega$ effect.

	As has been already mentioned in Sec.~\ref{sec:III}, in order for a large-scale magnetic-flux freezing to occur in a highly turbulent medium, we need mechanisms that balance or cancel the strong turbulent magnetic diffusivity effect. Without such balancing or transport suppression mechanisms, we only have Eq.~(\ref{eq:mean_B_eq_beta}), then no magnetic-flux freezing occurs. The cross-helicity effect, as well as the helicity or $\alpha$ effect, is one of such balancing mechanisms.
	
	In addition to the magnetic-flux freezing, in order to get a toroidal component of the magnetic field from the poloidal magnetic field, we need a differential rotation of the toroidal velocity. The toroidal velocity should be faster or slower in the lower-latitude region than in the higher-latitude region. Without such differential rotation profiles of the toroidal velocity, the toroidal magnetic field can not be generated from the poloidal magnetic field in the framework of the $\Omega$ effect. The toroidal rotation observed at the surface of the Sun, the prograde rotation in the equatorial or lower-latitude region, is considered to support the scenario of a dynamo that utilizes the $\Omega$ effect.
	
	During the process generating the toroidal magnetic field from the poloidal magnetic field by differential rotation, we in general need a magnetic-field reconnection, which makes a new configuration possible from the old one. Original poloidal field is distorted by the differential rotation. Since such distortion is most strong at the midplane or equatorial plane, it is most likely that the reconnection occurs there. Then we have a toroidal magnetic-field configuration whose direction is opposite with respect to the midplane.	

	These considerations immediately suggest that the $\Omega$ effect utilized for the toroidal magnetic-field generation from the poloidal one is based on a subtle combination of several requirements. It requires a particular type of differential rotation profile that is suitable for the toroidal field generation. It requires a particular spatial distribution of the effective magnetic diffusivity that is convenient for both the magnetic-flux freezing in some region and the magnetic reconnection in other region. In this sense, it is quite natural that model simulations that utilize the $\Omega$ effect are very sensitive to the variation of velocity profile including the incorporation of the meridional rotation and the value and spatial distribution of the effective magnetic diffusivity.

\subsubsection{\label{sec:III-B-2}Cross-helicity effect}

	If the cross-helicity effect is the main balancer for the turbulent magnetic diffusivity effect, the mean induction equation in a stationary state has a special solution as follows.
	
	We divide the mean magnetic field and electric-current density as
\begin{equation}
	{\bf{B}} = {\bf{B}}_0 + \delta {\bf{B}},\;\;
	{\bf{J}} = {\bf{J}}_0 + \delta {\bf{J}},
	\label{eq:mag_expand}
\end{equation}
where ${\bf{B}}_0$ and ${\bf{J}}_0 (= \nabla \times {\bf{B}}_0)$ are the mean magnetic field and electric-current density without the cross-helicity effect, while $\delta{\bf{B}}$ and $\delta{\bf{J}} (= \nabla \times \delta{\bf{B}})$ are the mean fields representing the first-order contribution through the cross-helicity effect. Substituting Eq.~(\ref{eq:mag_expand}) into Eq.~(\ref{eq:mean_B_eq_E_M_full}) with the $\alpha$-related term dropped, we have equations
\begin{equation}
	\frac{\partial {\bf{B}}_0}{\partial t}
	= \nabla \times \left( {{\bf{U}} \times {\bf{B}}_0} \right)
	- \nabla \times \left( {\beta \nabla \times {\bf{B}}_0} \right)
	\label{eq:B0_eq}
\end{equation}
for the zeroth-order field, and
\begin{equation}
	\frac{\partial \delta{\bf{B}}}{\partial t}
	= \nabla \times \left( {{\bf{U}} \times \delta{\bf{B}}} \right)
	- \nabla \times \left( {
		\beta \nabla \times \delta{\bf{B}}
		- \gamma \nabla \times {\bf{U}}
	} \right)
	\label{eq:delta_B_eq}
\end{equation}
for the first-order field.

\begin{figure}[htb]
\includegraphics[width=.35\textwidth]{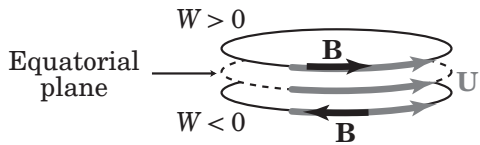}
\caption{\label{fig:toroidal_cross_hel} Toroidal magnetic-field generation due to the cross-helicity dynamo. An antisymmetric toroidal-field configuration is induced from a symmetric velocity profile coupled with an antisymmetric cross-helicity distribution. No differential rotation is required.}
\end{figure}

	Equation~(\ref{eq:delta_B_eq}) represents the magnetic induction $\delta {\bf{B}}$ arising from the turbulent cross helicity. Equation~(\ref{eq:delta_B_eq}) has a particular solution
\begin{equation}
	\delta {\bf{B}} 
	= \frac{\gamma}{\beta} {\bf{U}}
	= C_{\rm{W}} \frac{W}{K} {\bf{U}}
	\label{eq:delta_B_sol}
\end{equation}
for a stationary state. Here, $C_{\rm{W}} (= C_\gamma / C_\beta)$ is a model constant whose value is estimated as $O(10^{-1})-O(1)$ [See Eqs.~(\ref{eq:beta_model}) and (\ref{eq:gamma_model})]. Here, the proportional coefficient $W/K$ is the turbulent cross helicity scaled by the turbulent MHD energy. This scaled cross helicity is the most important quantity in the cross-helicity dynamo formulation. The value of $W/K$ is not positive-definite, but its absolute value is bounded as
\begin{equation}
	\frac{|W|}{K}
	= \frac{|\langle {{\bf{u}}' \cdot {\bf{b}}'} \rangle|}
		{\langle {{\bf{u}}'{}^2 + {\bf{b}}'{}^2} \rangle /2}
	\le 1.
	\label{eq:W/K_constraint}
\end{equation}
The cross-helicity dynamo solution [Eq.~(\ref{eq:delta_B_sol})] requires a finite value of turbulent cross helicity for it to work. In the absence of cross helicity in turbulence, Eq.~(\ref{eq:delta_B_sol}) gives no contribution at all for the magnetic-field generation. In this sense, the spatial and temporal distributions of the cross helicity should be intensively examined. In order to fully investigate such distributions, we have to solve the evolution equation of $W$ [Eq.~(\ref{eq:G_eq}) with Eq.~(\ref{eq:W_evol})]. However, without considering the full complications of the cross-helicity evolution, we can derive several consequences of Eq.~(\ref{eq:delta_B_sol}) as follows. The turbulent cross helicity is a pseudoscalar, which describes breakage of mirrorsymmetry. As has been referred to in Sec. I, cross helicity is considered to be a measure of the asymmetry with respect to the magnetic field. If a magnetic field threads through plasmas with the turbulent energy distributed symmetrically with respect to the midplane, we have asymmetry between the directions parallel and antiparallel to the magnetic field. Then, we see from Eq.~(\ref{eq:spatio_dis_W}) that the spatial distribution of the turbulent cross helicity is antisymmetric. If such an antisymmetric distribution of $W$ is coupled with the symmetric toroidal velocity, from Eq.~(\ref{eq:delta_B_sol}) we have antisymmetric toroidal magnetic field (Fig.~\ref{fig:toroidal_cross_hel}).

	A poloidal magnetic field coupled with the inhomogeneous turbulence leads to antisymmetric distribution of the cross helicity in turbulence [Eqs.~(\ref{eq:B_grad_K}) and (\ref{eq:spatio_dis_W})]. This turbulent cross helicity coupled with the toroidal velocity gives the toroidal magnetic field [Eq.~(\ref{eq:delta_B_sol})]. A prominent feature of this scenario is that it does not require a particular differential rotation of the toroidal velocity for the sake of toroidal magnetic-field generation. In the framework of cross-helicity dynamo, differential rotation is not directly related to the toroidal field generation. Differential rotation is important only to sustain the energy and cross helicity in turbulence. As we see in Eq.~(\ref{eq:W/K_constraint}), the magnitude of the turbulent cross helicity is bounded by that of the turbulent MHD energy. Unless we have enough amount of turbulent MHD energy, we will not have enough cross helicity in turbulence. Only in this sense, the differential rotation of plasmas is required for the dynamo process to work.

	Very recent numerical simulations showed that a dynamo model constituted of the evolution equations of the toroidal and poloidal magnetic fields as well as the turbulent cross helicity can reproduce an oscillatory butterfly diagram of the solar magnetic activity without resort to the $\Omega$ effect.\cite{pip2010} These results will be reported in the forthcoming paper.

\section{\label{sec:IV}Cross-helicity effects in magnetic reconnection}

	Magnetic reconnections are associated with large-scale plasma flows with some typical configurations. We focus our attention on such large-scale flow structures associating with the magnetic reconnection, and examine how turbulence influences such flow structures as well as how such flow structures determine the properties of turbulence.

\subsection{\label{sec:IV-A}Cross-helicity effects in momentum balance}

	So far we treated only the mean induction equation. Here we examine the cross-helicity effects in the equation of motions. For this purpose, we consider the mean vorticity equation
\begin{eqnarray}
	\frac{\partial {\bf{\Omega}}}{\partial t}
	&=& \nabla \times \left[ {
	\left( {{\bf{U}} - \frac{\gamma}{\beta} {\bf{B}}} \right) \times {\bf{\Omega}}
	+ \nu_{\rm{K}} \nabla^2 \left( {{\bf{U}} - \frac{\gamma}{\beta} {\bf{B}}} \right)
	} \right]
	\nonumber\\
	&+& \nabla \times \left[ { 
	{\bf{F}} + \frac{1}{\beta} 
	\left( {{\bf{U}} \times {\bf{B}}} \right) \times {\bf{B}}
	- \frac{1}{\beta} \frac{\partial {\bf{A}}}{\partial t} \times {\bf{B}}
	} \right],
	\label{eq:Omega_eq}
\end{eqnarray}	
which is a direct consequence of the expressions for the Reynolds stress $\mbox{\boldmath$\cal R$}$ [Eq. (\ref{eq:Re_str_exp})] and the turbulent electromotive force ${\bf{E}}_{\rm{M}}$ [Eq.~(\ref{eq:E_M_exp})] substituted into the mean velocity equation (${\bf{F}}$: mean external force, ${\bf{A}}$: mean vector potential). For the derivation of Eq.~(\ref{eq:Omega_eq}), see Appendix~\ref{app:B}.

	As we have seen in Eq.~(\ref{eq:G_eq}) with Eq.~(\ref{eq:W_evol}), it is difficult to sustain a finite cross helicity in turbulence in the absence of the mean magnetic field. In order to clearly understand the role of the turbulent cross helicity in momentum transport, we divide the mean velocity and vorticity as
\begin{equation}
	{\bf{U}} = {\bf{U}}_0 + \delta {\bf{U}},\;\;
	\mbox{\boldmath$\Omega$} 
	= \mbox{\boldmath$\Omega$}_0 + \delta\mbox{\boldmath$\Omega$}.
	\label{eq:U_expand}
\end{equation}
Here, ${\bf{U}}_0$ and $\mbox{\boldmath$\Omega$}_0$ are the mean velocity and vorticity without the mean magnetic-field effects, while $\delta {\bf{U}}$ and $\delta \mbox{\boldmath$\Omega$}$ are the mean fields representing the first-order effects of the mean magnetic field through the turbulent cross helicity. Substituting Eq.~(\ref{eq:U_expand}) into Eq.~(\ref{eq:Omega_eq}), we have equations
\begin{equation}
	\frac{\partial {\bf{\Omega}}_0}{\partial t}
	= \nabla \times \left( {
		{\bf{U}}_0 \times \mbox{\boldmath$\Omega$}_0
		+ \nu_{\rm{K}} \nabla^2 {\bf{U}_0}
		+ {\bf{F}}
	} \right)
	\label{eq:Omega_0_eq}
\end{equation}
for the zeroth-order field, and
\begin{equation}
	\frac{\partial \delta{\bf{\Omega}}}{\partial t}
	= \nabla \times \left[ {
		\left( {\delta{\bf{U}} - \frac{\gamma}{\beta} {\bf{B}}} \right)
		\times \mbox{\boldmath$\Omega$}_0
		+ \nu_{\rm{K}} \nabla^2 \left( {
		\delta{\bf{U}} - \frac{\gamma}{\beta} {\bf{B}}
		} \right)
	} \right]
	\label{eq:delta_Omega_eq}
\end{equation}
for the first-order field.

	Equation~(\ref{eq:Omega_0_eq}) expresses the evolution of the large-scale flow structure $\mbox{\boldmath$\Omega$}_0$, which is subject to the turbulent viscosity $\nu_{\rm{K}}$ and the external force ${\bf{F}}$. On the other hand, Eq.~(\ref{eq:delta_Omega_eq}) represents the variation of the large-scale flow structure $\delta \mbox{\boldmath$\Omega$}$ arising from the turbulent cross helicity or $\gamma$ effect. We see from Eq.~(\ref{eq:delta_Omega_eq}) that, for a given large-scale flow structure $\mbox{\boldmath$\Omega$}_0$,
\begin{equation}
	\delta {\bf{U}} 
	= \frac{\gamma}{\beta} {\bf{B}}
	= C_{\rm{W}} \frac{W}{K} {\bf{B}}
	\label{eq:delta_U_sol}
\end{equation}
is a particular solution of the stationary state. Equation~(\ref{eq:delta_U_sol}) suggests that we have a variation of momentum transfer that is proportional to the large-scale magnetic field multiplied by the normalized cross helicity.

	Here we should make some remarks on the validity of Eq.~(\ref{eq:delta_U_sol}) in the compressible case. In compressible flows, we have several other terms arising from the density variation in Eq.~(\ref{eq:Omega_eq}) or (\ref{eq:mean_Omega_eq_exp}). If we neglect the density fluctuation ($\rho' = 0$) and consider only the mean density stratification $\rho = \overline{\rho}$ ($\overline{\rho}$: mean density, $\rho'$: fluctuation density), one of the main contributions is the baroclinic effect in the mean vorticity equation:
\begin{equation}
	\nabla \times \left( {
	- \frac{1}{\overline{\rho}} \nabla P
	} \right)
	= \frac{\nabla \overline{\rho} \times \nabla P}{\overline{\rho}^2}.
	\label{eq:baroclinic}
\end{equation}
If we further assume that the cross helicity does not affect the mean density and pressure ($\overline{\rho} = \overline{\rho}_0$ and $P=P_0$ or $\delta \overline{\rho} = \delta P = 0$), Eq.~(\ref{eq:baroclinic}) is reduced to
\begin{equation}
	\nabla \times \left( {
	- \frac{1}{\overline{\rho}} \nabla P
	} \right)
	= \frac{\nabla \overline{\rho}_0 \times \nabla P_0}{\overline{\rho}_0^2},
	\label{eq:reduced_baroclinic}
\end{equation}
and no contribution to the first-order equation [Eq.~(\ref{eq:delta_Omega_eq})]. In such a case, the special solution [Eq.~(\ref{eq:delta_U_sol})] is obtainable even in the compressible case.

\subsection{\label{sec:IV-B}Reconnection environment viewed from the large-scale flow structure}

	It is pointed out that properties of magnetic reconnection such as the scaling law and maximum reconnection rate depend heavily on the form, nature, and value of the imposed boundary conditions.\cite{for1987,pri2000} Here we examine the reconnection environment from the viewpoint of large-scale structure. We consider the circumstances or global surroundings of the reconnection region are highly turbulent. Such turbulence is a consequence of the large-scale inhomogeneity of the mean fields such as the velocity strain, electric current, etc. In this sense, turbulence in magnetic reconnections here is driven by large-scale structures. At the same time, we also consider the self-generation mechanisms of turbulence in the magnetic-reconnection process.
	
\begin{figure}[htb]
\includegraphics[width=.30\textwidth]{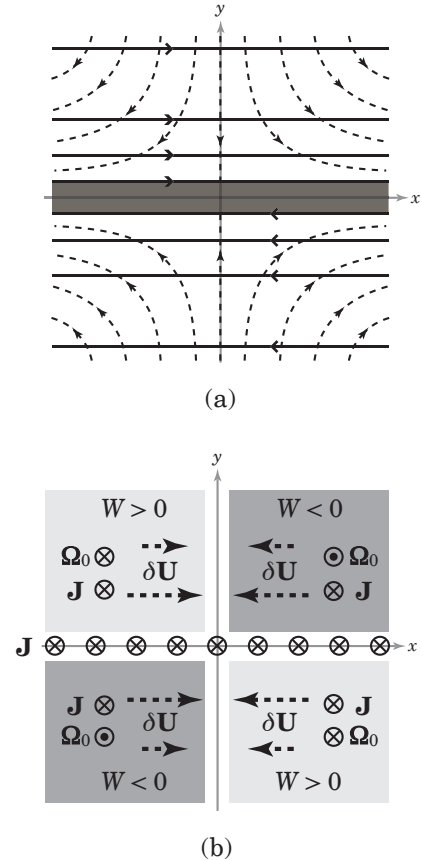}
\caption{\label{fig:flow_in_mr} Magnetic reconnection and turbulent cross helicity. (a) Magnetic-field lines and streamlines for a large-scale inflow configuration. Solid lines: magnetic field, Dashed lines: streamline. Shaded region represents the reconnection or diffusion region. (b) Directions of the mean fields and induced velocity due to the turbulent cross helicity. Directions of the mean electric-current density ${\bf{J}}$ and the mean vorticity $\mbox{\boldmath$\Omega$}_0$ are shown with the sign of the turbulent cross helicity $W$. Directions of the induced velocity due to the cross-helicity effect [Eq.~(\ref{eq:delta_U_sol})] are also indicated by the dashed arrows.}
\end{figure}

	Since the argument leading to the result [Eq.~(\ref{eq:delta_U_sol})] is quite general, we can consider any general situations including the magnetic reconnection in three dimensions. For the sake of simplicity of description, here we consider a simple situation where a current sheet is associated with a large-scale inflow as in Fig.~\ref{fig:flow_in_mr}(a).

	Recalling the cross-helicity generation mechanism discussed in Sec.~\ref{sec:II-C}, we see that the generation mechanism related to the turbulent electromotive force [Eq.~(\ref{eq:Pw_Em})] may be most important in this flow configuration. As we see in Fig.~\ref{fig:flow_in_mr}(b), the large-scale vorticity $\mbox{\boldmath$\Omega$}_0$ is spatially distributed antisymmetrically with respect to the midplane of current sheet ($x$ axis, plane with $y=0$), and also with respect to the center plane ($y$ axis, plane with $x=0$). Since the large-scale electric-current density ${\bf{J}}$ is aligned with the negative $z$ direction everywhere in this case, the sign of the product of $\mbox{\boldmath$\Omega$}_0$ and ${\bf{J}}$, ${\bf{J}} \cdot \mbox{\boldmath$\Omega$}_0$, is distributed as
\begin{equation}
	\left\{ {
	\begin{array}{ll}
		{\bf{J}} \cdot \mbox{\boldmath$\Omega$}_0 < 0 &\mbox{for}\;\; xy>0,
		\rule{0.ex}{0.ex}\\
		{\bf{J}} \cdot \mbox{\boldmath$\Omega$}_0 > 0 &\mbox{for}\;\; xy<0.
		\rule{0.ex}{3.ex}
	\end{array}
	} \right.
	\label{eq:J_Omega_mr}
\end{equation}
Consequently, the production of the turbulent cross helicity is positive and negative in the regions with $xy<0$ and $xy>0$, respectively. Then it is highly probable that the turbulent cross helicity is spatially distributed as
\begin{equation}
	\left\{ {
	\begin{array}{ll}
		W < 0 &\mbox{for}\;\; xy>0,
		\rule{0.ex}{0.ex}\\
		W > 0 &\mbox{for}\;\; xy<0
		\rule{0.ex}{3.ex}
	\end{array}
	} \right.
	\label{eq:W_distr_mr}
\end{equation}
as depicted in Fig.~\ref{fig:flow_in_mr}(b).

	Outside of the current sheet region, it follows from Eq.~(\ref{eq:delta_U_sol}) that the velocity variation due to the cross-helicity effect is expressed as
\begin{equation}
	\everymath{\displaystyle}
	\delta {\bf{U}}
	= \left\{ {
	\begin{array}{ll}
		- \frac{|\gamma|}{\beta} {\bf{B}}
		= - C_{\rm{W}} \frac{|W|}{K} {\bf{B}} &\mbox{for}\;\; x>0,
		\rule{0.ex}{0.ex}\\
		\frac{|\gamma|}{\beta} {\bf{B}}
		= C_{\rm{W}} \frac{|W|}{K} {\bf{B}} &\mbox{for}\;\; x<0,
		\rule{0.ex}{5.ex}
	\end{array}
	} \right.
	\label{eq:delta_U_field}
\end{equation}

	From the symmetry of the plasma flow, the large-scale vorticity vanishes on the midplane ($y=0$) and center plane ($x=0$), so we have no cross-helicity-generation mechanism there. The turbulent cross helicity changes its sign across the current sheet (midplane) and the center plane. For instance, in the positive $x$ region in Fig.~\ref{fig:flow_in_mr}, the sign of $W$ changes from positive ($y>0$) to negative ($y<0$) across the current sheet. As this consequence, the magnitude of the cross helicity is small in the vicinity of the current sheet. Equation~(\ref{eq:delta_U_field}) thus makes a very small contribution to the velocity in the current-sheet region (and also in the center-plane region).
	
	In the absence of the turbulent cross helicity ($W = \gamma = 0$), we have no induced flows, then the flow configuration remains as it was originally set (${\bf{U}}_0$ and $\mbox{\boldmath$\Omega$}_0$). If the flow is a potential one, we have no mean vorticity. In this situation, we have no cross-helicity generation since ${\bf{J}} \cdot \mbox{\boldmath$\Omega$}_0 = 0$ everywhere. A stagnation-point-like flow is such an example. On the other hand, in the presence of the turbulent cross helicity, the induced flow due to the cross-helicity effect should alter the flow profile. How much flow configurations are modulated depends on $\gamma / \beta$ or $W / K$, the turbulent cross helicity scaled by the turbulent MHD energy. For example, in the case of maximum cross helicity ($|W| / K = 1$), the induced flow speed is close to the Alfv\'{e}n speed. As this result, outside the current-sheet region, the inflow is converging toward the central plane ($x=0$), and in the current-sheet region the outflow is concentrated very narrow region in the vicinity of the midplane ($y=0$), where the magnitude of $W$ is very small.
	
\begin{figure}[htb]
\includegraphics[width=.35\textwidth]{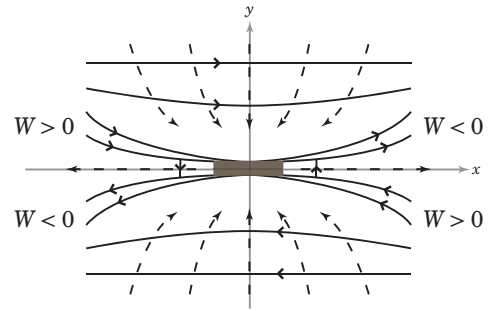}
\caption{\label{fig:flow_in_fast_mr} Flow and magnetic-field configurations in fast magnetic reconnection. Solid lines: magnetic field, Dashed lines: plasma flow. Shaded region represents the reconnection or diffusion region. Signs of the turbulent cross helicity $W$ associated with the slow shocks are also indicated. Sign of $W$ is negative (or positive) for the region where the Alfv\'{e}n wave is propagating in the directions parallel (or antiparallel) to the large-scale magnetic field.}
\end{figure}

	In the mean magnetic equation, the turbulent cross helicity coupled with the large-scale vortical motion gives rise to the induced magnetic field that is aligned with the mean velocity [Eq.~(\ref{eq:delta_B_sol})]. The spatial distribution of the turbulent cross helicity, Eq.~(\ref{eq:W_distr_mr}), leads to the induced field:
\begin{equation}
	\everymath{\displaystyle}
	\delta {\bf{B}} 
	= \left\{ { 
	\begin{array}{ll}
	- \frac{|\gamma|}{\beta} {\bf{U}}
	= - C_{\rm{W}} \frac{|W|}{K} {\bf{U}}&\mbox{for}\;\; xy>0,\rule{0.ex}{0.ex}\\
	\frac{|\gamma|}{\beta} {\bf{U}}
	= C_{\rm{W}} \frac{|W|}{K} {\bf{U}} &\mbox{for}\;\; xy<0.
	\rule{0.ex}{5.ex}\end{array}
	} \right.
	\label{eq:delta_B_in_mr}
\end{equation}
Consequently, the magnetic field
\begin{equation}
	{\bf{B}} = {\bf{B}}_0 + \delta {\bf{B}}
	\label{eq:toal_B_in_mr}
\end{equation}
turns to be a curved field configuration as schematically depicted in Fig.~\ref{fig:flow_in_fast_mr}. Combination of the inflow converging toward the $y$ axis with the curved magnetic field line gives a X-point-like configuration, which is known to be suitable for the fast reconnection. In the absence of the turbulent cross helicity, such modulations of the large-scale velocity and magnetic fields never occur. This suggests that the presence of cross helicity with a favorable mean-field configuration leads to the enhancement of magnetic reconnection rate as compared with the case without cross helicity.
	
\subsection{\label{sec:IV-C}Reconnection rate}

	As we have already mentioned, the width of the diffusion region in magnetic reconnection is reduced because of the converging flow due to Eq.~(\ref{eq:delta_U_field}) and the curvature of the magnetic field due to Eq.~(\ref{eq:delta_B_in_mr}), which would lead to the enhancement of the reconnection rate. Here, we estimate the reconnection rate with the aid of the expressions for the cross-helicity effects in the magnetic induction [Eq.~(\ref{eq:delta_B_sol})] and in the momentum balance [Eq.~(\ref{eq:delta_U_sol})].

	As the dimensionless reconnection rate, we adopt the inflow Alfv\'{e}n Mach number\cite{pri2000}
\begin{equation}
	M_{\rm{in}} \equiv \frac{U_{\rm{in}}}{V_{\rm{Ain}}}
	= \frac{U_{\rm{in}}}{B_{\rm{in}}},
	\label{eq:inflow_alfven_mach_def}
\end{equation}
where $B_{\rm{in}}$ is the magnetic field measured in the Alfv\'{e}n speed unit ($B_{\rm{in}} \equiv V_{\rm{Ain}} = B_{\rm{in}}^\ast /\sqrt{\mu \rho}$, $V_{\rm{Ain}}$: Alfv\'{e}n speed, $B_{\rm{in}}^\ast$: magnetic field measured in the physical unit).

\begin{figure}[htb]
\includegraphics[width=.45\textwidth]{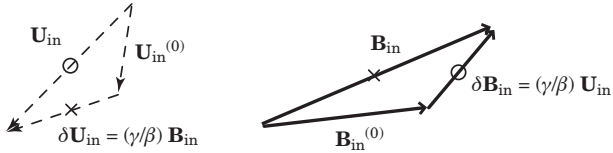}
\caption{\label{fig:field_modulation} Modulation of the velocity (Left) and magnetic field (Right) due to the cross-helicity effects. In this figure, the case with the negative cross helicity ($W<0$, $\gamma < 0$) is drawn.}
\end{figure}

	Let us denote the inflow velocity and magnetic field in the absence of the cross-helicity effects as ${\bf{U}}_{\rm{in}}^{(0)}$ and ${\bf{B}}_{\rm{in}}^{(0)}$, respectively.  From Eqs.~(\ref{eq:delta_U_sol}) and (\ref{eq:delta_B_sol}), the inflow velocity and magnetic field with the cross-helicity effects, ${\bf{U}}_{\rm{in}}$ and ${\bf{B}}_{\rm{in}}$, are expressed as
\begin{equation}
	{\bf{U}}_{\rm{in}} = {\bf{U}}_{\rm{in}}^{(0)} + \delta {\bf{U}}_{\rm{in}}
	= {\bf{U}}_{\rm{in}}^{(0)} + \frac{\gamma}{\beta} {\bf{B}}_{\rm{in}},
	\label{eq:modulated_U_in}
\end{equation}
\begin{equation}
	{\bf{B}}_{\rm{in}} = {\bf{B}}_{\rm{in}}^{(0)} + \delta {\bf{B}}_{\rm{in}}
	= {\bf{B}}_{\rm{in}}^{(0)} + \frac{\gamma}{\beta} {\bf{U}}_{\rm{in}},
	\label{eq:modulated_B_in}
\end{equation}
respectively (Fig.~\ref{fig:field_modulation}). This system of equations are written in a matrix form as
\begin{equation}
	\begin{pmatrix}
		1 & - \frac{\gamma}{\beta}\\ 
		- \frac{\gamma}{\beta} & 1 
	\end{pmatrix}
	\begin{pmatrix}
		{}^t{\bf{U}}_{\rm{in}}\\ 
		{}^t{\bf{B}}_{\rm{in}}
	\end{pmatrix}
	= \begin{pmatrix}
		{}^t{\bf{U}}_{\rm{in}}^{(0)}\\ 
		{}^t{\bf{B}}_{\rm{in}}^{(0)}
	\end{pmatrix},
	\label{eq:U_in0_B_in0}
\end{equation}
where ${}^t{\bf{U}}_{\rm{in}} = (U_{\rm{in}}^x, U_{\rm{in}}^y)$, ${}^t{\bf{B}}_{\rm{in}} = (B_{\rm{in}}^x, B_{\rm{in}}^y)$, etc. Solving Eq.~(\ref{eq:U_in0_B_in0}) with respect to $^t{\bf{U}}_{\rm{in}}$ and $^t{\bf{B}}_{\rm{in}}$, we have
\begin{equation}
	\begin{pmatrix}
		{}^t{\bf{U}}_{\rm{in}}\\ 
		{}^t{\bf{B}}_{\rm{in}}
	\end{pmatrix}
	= \left[ {1 - \left( {\frac{\gamma}{\beta}} \right)^2} \right]^{-1}
	\begin{pmatrix}
		1 & \frac{\gamma}{\beta}\\ 
		\frac{\gamma}{\beta} & 1 
	\end{pmatrix}
	\begin{pmatrix}
		{}^t{\bf{U}}_{\rm{in}}^{(0)}\\ 
		{}^t{\bf{B}}_{\rm{in}}^{(0)}
	\end{pmatrix}.
	\label{eq:U_in_B_in}
\end{equation}
Then, for given ${\bf{U}}_{\rm{in}}^{(0)}$ and ${\bf{B}}_{\rm{in}}^{(0)}$, we can calculate ${\bf{U}}_{\rm{in}}$ and ${\bf{B}}_{\rm{in}}$.

	From Eq.~(\ref{eq:U_in_B_in}), the magnitude of the inflow velocity and magnetic field are given as
\begin{eqnarray}
	\lefteqn{
	U_{\rm{in}} \equiv |{\bf{U}}_{\rm{in}}|
	= \left[ {1 - \left( {\frac{\gamma}{\beta}} \right)^2 } \right]^{-1/2} \times
	}\nonumber\\
	& & \hspace{20pt} \left[ {
		{{\bf{U}}_{\rm{in}}^{(0)}}^2 
		+  \left( { \frac{\gamma}{\beta}} \right)^2 {{\bf{B}}_{\rm{in}}^{(0)}}^2 
		+ 2 \frac{\gamma}{\beta} 
			{\bf{U}}_{\rm{in}}^{(0)} \cdot {\bf{B}}_{\rm{in}}^{(0)}
	}\right]^{1/2}\hspace{-12pt},
	\label{eq:U_in_magnitude}
\end{eqnarray}
\begin{eqnarray}
	\lefteqn{
	B_{\rm{in}} \equiv |{\bf{B}}_{\rm{in}}|
	= \left[ {1 - \left( {\frac{\gamma}{\beta}} \right)^2 } \right]^{-1/2} \times
	}\nonumber\\
	& & \hspace{20pt} \left[ {
		\left( { \frac{\gamma}{\beta}} \right)^2 {{\bf{U}}_{\rm{in}}^{(0)}}^2 
		+  {{\bf{B}}_{\rm{in}}^{(0)}}^2 
		+ 2 \frac{\gamma}{\beta} 
			{\bf{U}}_{\rm{in}}^{(0)} \cdot {\bf{B}}_{\rm{in}}^{(0)}
	}\right]^{1/2}\hspace{-12pt}.
	\label{eq:B_in_magnitude}
\end{eqnarray}

	Then, the reconnection rate is expressed as
\begin{equation}
	M_{\rm{in}} = \left[ {
		\frac{{M_{\rm{in}}^{(0)}}^2 + (\gamma/\beta)^2 
			+ (2\gamma / \beta) M_{\rm{in}}^{(0)} \cos \theta_0}
			{(\gamma/\beta)^2 {M_{\rm{in}}^{(0)}}^2 + 1 
			+  (2\gamma / \beta) M_{\rm{in}}^{(0)} \cos \theta_0}
	} \right]^{1/2}\hspace{-12pt},
	\label{eq:Min_exp}
\end{equation}
where $M_{\rm{in}}^{(0)} = U_{\rm{in}}^{(0)} / B_{\rm{in}}$ is the reconnection rate in the absence of the cross-helicity effect, and $\theta_0$ is the angle between ${\bf{U}}_{\rm{in}}^{(0)}$ and ${\bf{B}}_{\rm{in}}^{(0)}$. In the case of converging inflow and X-point like curved magnetic field such as in Fig.~\ref{fig:flow_in_fast_mr}, the sign of $\cos\theta_0$ or ${\bf{U}}_{\rm{in}}^{(0)} \cdot {\bf{B}}_{\rm{in}}^{(0)}$ obeys the quadrupole distribution similar to that of $W$ [Eq.~(\ref{eq:W_distr_mr})]:
\begin{equation}
		\left\{ {
	\begin{array}{ll}
		\cos \theta_0 < 0 &\mbox{for}\;\; xy>0,
		\rule{0.ex}{0.ex}\\
		\cos \theta_0 > 0 &\mbox{for}\;\; xy<0.
		\rule{0.ex}{3.ex}
	\end{array}
	} \right.
	\label{eq:cos_theta_distr}
\end{equation}
Then, the sign of $(2\gamma/\beta) M_{\rm{in}}^{(0)} \cos\theta_0$ in Eq.~(\ref{eq:Min_exp}), as well as the $(\gamma/\beta)^2$-related contributions, is always positive. In this sense, in the converging inflow cases it is not the $\gamma/\beta$ and $\cos\theta_0$ themselves but the magnitudes of them which determine the modulation of $M_{\rm{in}}$.
	                                       
	 If ${\bf{U}}_{\rm{in}}^{(0)}$ is perpendicular to ${\bf{B}}_{\rm{in}}^{(0)}$ ($\theta_0 = \pi/2$), Eq.~(\ref{eq:Min_exp}) is reduced to
\begin{equation}
	M_{\rm{in}}(\theta_0 = \pi/2) = \left[ {
		\frac{{M_{\rm{in}}^{(0)}}^2 + (\gamma/\beta)^2}
			{(\gamma/\beta)^2 {M_{\rm{in}}^{(0)}}^2 + 1}
	} \right]^{1/2}\hspace{-12pt}.
	\label{eq:Min_exp_pi/2}
\end{equation}

\begin{figure}[htb]
\includegraphics[width=.45\textwidth]{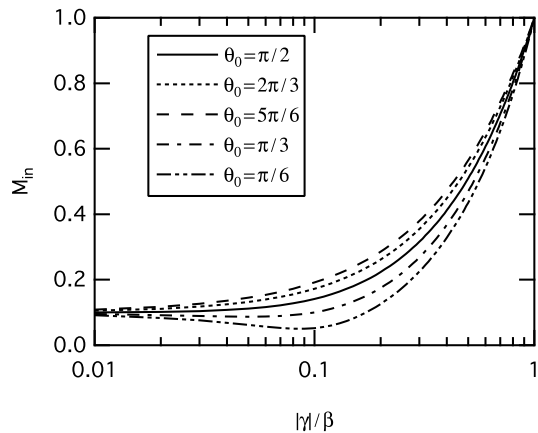}
\caption{\label{fig:reconnection_rates} Variation of the reconnection rate $M_{\rm{in}}$ against the scaled turbulent cross helicity $\gamma / \beta$. The angle between the inflow velocity and magnetic field in the absence of the cross-helicity effects, ${\bf{U}}_{\rm{in}}^{(1)}$ and ${\bf{B}}_{\rm{in}}^{(0)}$, is denoted by $\theta_0$. ------: $\theta_0 = \pi/2$; $\cdots\cdots$: $\theta_0 = 2\pi/3$;  -~-~-: $\theta_0 = 5\pi/6$; --~$\cdot$~--, $\theta_0 = \pi/3$; --~$\cdot\cdot$~--, $\theta_0 = \pi/6$. In all plots, the magnetic reconnection rate without the cross-helicity effects, $M_{\rm{in}}^{(0)}$, is set equal to $0.1$.}
\end{figure}

	The behaviors of $M_{\rm{in}}$ with respect to $|\gamma|/\beta$ with several values of $\theta_0$ in the case of $M_{\rm{in}}^{(0)} = 0.1$ are plotted in Fig.~\ref{fig:reconnection_rates}. Here, as an example, we consider the domain with $xy>0$, where $\gamma <0$. As expected, we have no enhancement of $M_{\rm{in}}$ by the cross-helicity effect if $\gamma / \beta =0$. With increase in $|\gamma| / \beta$, $M_{\rm{in}}$ increases up to $1$ (at $|\gamma| / \beta =1$). Here, we should note that $\gamma / \beta$ is expressed by the turbulent cross helicity scaled by the turbulent MHD energy as
\begin{equation}
	\frac{|\gamma|}{\beta} = C_{\rm{W}} \frac{|W|}{K} < 1
	\label{eq:gamma/beta_W/K}
\end{equation}
with $1 > C_{\rm{W}} = O(10^{-1})-O(1)$, which corresponds to the fact that $C_\gamma$ [Eq.~(\ref{eq:gamma_model})] is smaller than $C_\beta$ [Eq.~(\ref{eq:beta_model})] but the orders of them are the same.

	These results suggest that the magnetic reconnection rate may be substantially enhanced by the cross-helicity effects if the magnitude of  the scaled turbulent cross helicity is large enough such as
\begin{equation}
	\frac{|W|}{K} \gtrsim 0.1.
	\label{eq:high_crss_hel}
\end{equation}
On the other hand, the cross-helicity effect may be negligible for the magnetic reconnection rate in the case of the turbulent cross helicity of
\begin{equation}
	\frac{|W|}{K} \ll 0.1.
	\label{eq:low_crss_hel}
\end{equation}

	In the case of non-converging inflow geometry, the signs of $W$ and $\cos\theta_0$ may be opposite to each other, then the sign of $(2\gamma/\beta) M_{\rm{in}}^{(0)} \cos\theta_0$ is negative. In such a case, the cross-helicity effects work for reducing the reconnection rate for small $|\gamma|/\beta$-value domains since the contribution of $(2\gamma/\beta) M_{\rm{in}}^{(0)} \cos\theta_0$ is more dominant than that of $(\gamma / \beta)^2$. In Fig.~\ref{fig:reconnection_rates}, the $M_{\rm{in}}$ behavior in such situations ($\theta_0 = \pi/6, \pi/3$) are also plotted.

	Another important point to remark is that the turbulent cross helicity distributions preferable to the reconnection are expected to be sustained in the down-stream region or at the after-reconnection stage. If there is an asymmetry between the directions parallel and antiparallel to the magnetic field, we have a certain production mechanism of the non-vanishing turbulent cross helicity. As is well known, X-point reconnection and slow shock regions are source of the Alfv\'{e}n waves. For instance, the anisotropic ion beams in the plasma sheet boundary layer, which is the so-called PSBL ions in the magnetotail,~\cite{gri2011} is known to excite the Alfv\'{e}n waves by the ion cyclotron beam instability.~\cite{hos1998} The reflected ion beams from the slow shock front may excite the Alfv\'{e}n wave in the shock upstream along the magnetic-field line as well.\cite{omi1992} Since the Alfv\'{e}n waves generated by slow  shocks propagate in the outward direction, the sign of the cross helicity associated with the Alfv\'{e}n waves is negative (or positive) if the large-scale magnetic field is parallel (or antiparallel) to the direction of the wave propagation. In terms of turbulent statistical quantities, this mechanism is related to the cross-helicity generation due to the inhomogeneity along the mean magnetic field, ${\bf{B}} \cdot \nabla K$ [Eq.~(\ref{eq:B_grad_K})]. The resultant cross-helicity distribution is consistent with the one expressed by Eq.~(\ref{eq:W_distr_mr}) as shown in Fig.~\ref{fig:flow_in_fast_mr}.
	
\begin{figure}[htb]
\includegraphics[width=.25\textwidth]{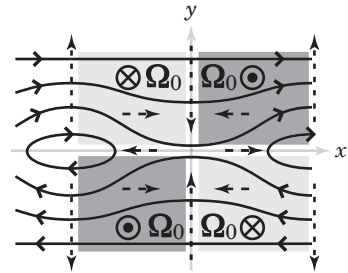}
\caption{\label{fig:tearing_mode} Tearing-mode configuration and distribution of the turbulent cross helicity.}
\end{figure}

	In relation to the spatial distribution of the turbulent cross helicity, we should note that the quadrupole symmetric distribution of the turbulent cross helicity [Eq.~(\ref{eq:W_distr_mr})] may be commonly or ubiquitously observed in the magnetic-reconnection configurations. For example, in the tearing-mode configuration, the large-scale vorticity $\mbox{\boldmath$\Omega$}_0$ associated with the plasma flow is distributed as depicted in Fig.~\ref{fig:tearing_mode}. Then the production of the turbulent cross helicity due to the coupling of $\bf{J}$ and $\mbox{\boldmath$\Omega$}_0$ is expressed as in Eq.~(\ref{eq:J_Omega_mr}), leading to the turbulent cross helicity spatially distributed as in Eq.~(\ref{eq:W_distr_mr}). These points should be carefully scrutinized using numerical simulations.

\section{\label{sec:V}Concluding remarks}

	Magnetic reconnection was viewed from the flow--turbulence interaction mediated by the cross-helicity effects. In addition to the enhancement of transport, the primary effect of turbulence, turbulent transport may be effectively suppressed by some other effects of turbulence. The {\it structure} information of turbulence represented by some pseudoscalar statistical quantities such as the turbulent cross, kinetic, current, magnetic helicities is the key for analyzing the transport suppression or structure formation due to turbulence, whereas the {\it intensity} information of turbulence is essential for describing the transport enhancement or structure destruction. In order to investigate the balance between these two effects, we first considered the cross-helicity production mechanisms in turbulence. The cross-helicity effects, which appear in the Reynolds stress [Eq.~(\ref{eq:Re_str_exp})] and the turbulent electromotive force [Eq.~(\ref{eq:E_M_exp})], were examined in the context of the magnetic-flux freezing in turbulence, turbulent dynamo process, and magnetic reconnection. In the arguments on magnetic reconnections, we stressed the importance of the large-scale flow structure and global magnetic-field configuration since these inhomogeneous structures constitute the conditions for the cross-helicity generation. The spatiotemporal distribution of the cross helicity, in turn, determines the transport properties of turbulence, then affects the magnetic reconnection rates. We showed that the large-scale flow and magnetic-field structures associated with the magnetic reconnection are favorable for the cross-helicity generation in some cases, and that the flow and magnetic-field structures induced by the cross-helicity effects are consistently favorable for the enhanced magnetic reconnections. In the absence of the turbulent cross helicity, these mutual interactions between the large-scale structure and turbulence will never appear at all in the magnetic reconnection process.

	This theoretical scenario of magnetic reconnection based on the mutual interaction between the large-scale structures and the turbulent cross helicity should be validated using numerical simulations. A few points should be remarked on such simulations. First, the theoretical formulation based on the cross-helicity effects through Eqs.~(\ref{eq:delta_B_sol}) and (\ref{eq:delta_U_sol}) extended in this paper is very general, and not limited to the two-dimensional cases. Estimate of reconnection-rate modulation and numerical validation of it in three dimensions are desired much especially in the context of astrophysical magnetic reconnection. However, we can start with two-dimensional cases since, unlike the kinetic and magnetic helicities, the cross helicity can exist even in the two dimensional cases. Secondly, boundary conditions for the numerical simulations may be important but in a limited sense. Actually, it was reported that the conditions on flows are important in determining several features of magnetic reconnection, including the nature of the inflow, the length of the diffusion region, etc.\cite{pri2000} However, in the scenario presented in the present paper, such boundary conditions are important only in the sense that they affect the spatial distribution of the turbulent cross helicity (more generally, distributions of relevant turbulent statistical quantities) through which properties of turbulent transport are determined. Numerical simulations validating the present theoretical idea are being prepared, and results will be reported in the forthcoming papers.

\begin{acknowledgments}
	One of the authors (NY) would like to cordially thank Alexander G. Kosovichev for his encouragement, without which NY might have not even started considering the turbulent effects in magnetic reconnections. Part of this work was performed during NY's stay at the Center for Turbulence Research (CTR), Stanford University/NASA Ames for the CTR Summer Program 2010 (June-July, 2010) and at the Nordic Institute for Theoretical Physics (NORDITA) as a guest researcher (February and July-August, 2011).
\end{acknowledgments}

\appendix

\section{\label{app:A}Cross-helicity dynamo and its solution}

\subsection{\label{app:A-1}Turbulent electromotive force}
	
	From a closure theory for inhomogeneous turbulence applied to magnetohydrodynamic (MHD) turbulence, the Reynolds (and turbulent Maxwell) stress $\mbox{\boldmath$\cal{R}$}$ and the turbulent electromotive force ${\bf{E}}_{\rm{M}}$ are known to be expressed as\cite{yos1990}
\begin{equation}
	{\cal{R}}^{\alpha\beta}
	= \frac{2}{3} K_{\rm{R}}
	- \nu_{\rm{K}} {\cal{S}}^{\alpha\beta}
	+ \nu_{\rm{M}} {\cal{M}}^{\alpha\beta},
	\label{eq:Re_strs_tsdia}
\end{equation}
\begin{equation}
	{\bf{E}}_{\rm{M}}
	= - \beta {\bf{J}} 
	+ \gamma \mbox{\boldmath$\Omega$}
	+ \alpha {\bf{B}},
	\label{eq:E_M_tsdia}
\end{equation}
where $K_R (\equiv \langle {{\bf{u}}'{}^2 - {\bf{b}}'{}^2} \rangle /2)$ is the residual energy, and $\mbox{\boldmath${\cal{S}}$}$ and $\mbox{\boldmath${\cal{M}}$}$ are the strain rates of the mean velocity and magnetic fields defined by
\begin{equation}
	{\cal{S}}^{\alpha\beta}
	= \frac{\partial U^\beta}{\partial x^\alpha}
	+ \frac{\partial U^\alpha}{\partial x^\beta},
	\label{eq:vel_strain}
\end{equation}
\begin{equation}
	{\cal{M}}^{\alpha\beta}
	= \frac{\partial B^\beta}{\partial x^\alpha}
	+ \frac{\partial B^\alpha}{\partial x^\beta}.
	\label{eq:mag_strain}
\end{equation}
In Eqs.~(\ref{eq:Re_strs_tsdia}) and (\ref{eq:E_M_tsdia}), $\nu_{\rm{K}}$, $\nu_{\rm{M}}$, $\beta$, $\gamma$, and $\alpha$ are the transport coefficients, which are known to be expressed as
\begin{eqnarray}
	\alpha &=& \frac{1}{3}
	\int d{\bf{k}} \int_{-\infty}^\tau \!\!\!d\tau_1
	\tilde{G}(k,{\bf{x}};\tau,\tau_1,t) \times \nonumber\\
	& & \times \left[ {
		-H_{uu}(k,{\bf{x}};\tau,\tau_1,t) + H_{bb}(k,{\bf{x}};\tau,\tau_1,t)
	} \right],
	\label{eq:alpha_spectral}
\end{eqnarray}
\begin{eqnarray}
	\beta &=& \frac{1}{3}
	\int d{\bf{k}} \int_{-\infty}^\tau \!\!\!d\tau_1
	\tilde{G}(k,{\bf{x}};\tau,\tau_1,t) \times \nonumber\\
 	& & \times \left[ {
		Q_{uu}(k,{\bf{x}};\tau,\tau_1,t) + Q_{bb}(k,{\bf{x}};\tau,\tau_1,t)
	} \right],
	\label{eq:beta_spectral}
\end{eqnarray}
\begin{eqnarray}
	\gamma &=& \frac{1}{3}
	\int d{\bf{k}} \int_{-\infty}^\tau \!\!\!d\tau_1
	\tilde{G}(k,{\bf{x}};\tau,\tau_1,t) \times \nonumber\\
 	& & \times \left[ {
	Q_{ub}(k,{\bf{x}};\tau,\tau_1,t) + Q_{bu}(k,{\bf{x}};\tau,\tau_1,t)
	} \right],
	\label{eq:gamma_spectral}
\end{eqnarray}
\begin{equation}
	\nu_{\rm{K}} = (7/5) \beta,\;\;
	\nu_{\rm{M}} = (7/5) \gamma.
	\label{eq:nu_K_beta_rel}
\end{equation}
Here, $\tilde{G}$ is the Green's function of MHD turbulence, and $H_{uu}$, $H_{bb}$, $Q_{uu}$, $Q_{ub}$, $Q_{ub}$ are the spectral functions of the turbulent kinetic helicity, current helicity, kinetic energy, magnetic energy, and cross helicity. Equations~(\ref{eq:alpha_spectral})-(\ref{eq:gamma_spectral}) show that $\alpha$, $\beta$, and $\gamma$ are determined by the spectral distributions of the residual helicity, energy, and cross helicity, respectively. These coefficients have spatiotemporally non-local nature through the Green's function.

	The simplest possible models for the transport coefficients $\alpha$, $\beta$, and $\gamma$ are given by using turbulent one-point quantities: the turbulent residual helicity $H(\equiv \langle {- {\bf{u}}' \cdot \mbox{\boldmath$\omega$}' + {\bf{b}}' \cdot {\bf{j}}'} \rangle)$, the turbulent MHD energy $K(\equiv \langle {{\bf{u}}'{}^2 + {\bf{b}}'{}^2} \rangle/2)$, and the turbulent cross helicity $W(\equiv \langle {{\bf{u}}' \cdot {\bf{b}}'} \rangle)$ respectively:
\begin{equation}
	\alpha = C_\alpha \tau H,
	\label{eq:alpha_model}
\end{equation}
\begin{equation}
	\beta = C_\beta \tau K,
	\label{eq:beta_model}
\end{equation}
\begin{equation}
	\gamma = C_\gamma \tau W
	\label{eq:gamma_model}
\end{equation}
($C_\alpha$, $C_\beta$, and $C_\gamma$: model constants). Here $\tau$ is the time scale of turbulence, which is often expressed by using the turbulent MHD energy scaled by its dissipation rate:
\begin{equation}
	\tau = K/\varepsilon.
	\label{eq:time_model}
\end{equation}

\subsection{\label{app:A-2}Cross-helicity dynamo solution}

	If we drop the $\alpha$- or helicity-related term in the turbulent electromotive force ${\bf{E}}_{\rm{M}}$ [Eq.~(\ref{eq:E_M_tsdia})] as
\begin{equation}
	{\bf{E}}_{\rm{M}} 
	= - \beta {\bf{J}} + \gamma \mbox{\boldmath$\Omega$},
	\label{eq:beta_gamma_E_M}
\end{equation}
the mean induction equation (\ref{eq:mean_B_eq_E_M_full}) yields
\begin{equation}
	\frac{\partial{\bf{B}}}{\partial t}
	= \nabla \times \left( {{\bf{U}} \times {\bf{B}}} \right)
	+ \nabla \times \left( {
		- \beta \nabla \times {\bf{B}}
		+ \gamma \nabla \times {\bf{U}}
	} \right),
	\label{eq:mean_B_eq_cross_helicity}
\end{equation}
where the molecular magnetic diffusivity $\eta$ have been dropped as compared with the turbulent counterpart $\beta$. In a stationary state, the mean magnetic field
\begin{equation}
	{\bf{B}} = \frac{\gamma}{\beta} {\bf{U}}
	= C_{\rm{W}} \frac{W}{K} {\bf{U}}
	\label{eq:cross_helicity_dynamo_sol}
\end{equation}
is a special solution of Eq.~(\ref{eq:mean_B_eq_cross_helicity}) provided that the spatial variation of $\beta / \gamma (= C_{\rm{W}} W/K)$ is not so large.\cite{yos1993}

\section{\label{app:B}Vorticity equation in the presence of the turbulent cross-helicity effects}

	The mean momentum equation in the rotational form is written as
\begin{eqnarray}
	\frac{\partial {\bf{U}}}{\partial t}
	&=& {\bf{U}} \times {\bf{\Omega}}
	+ {\bf{J} \times {\bf{B}}}
	- \nabla \cdot \mbox{\boldmath$\cal R$}
	+ {\bf{F}}\nonumber\\
	& & - \nabla \left( {
		P + \frac{1}{2} {\bf{U}}^2 
		+ \left\langle {\frac{1}{2} {\bf{b}}'{}^2} \right\rangle
	} \right),
	\label{eq:mean_U_eq_rot}
\end{eqnarray}
where $\mbox{\boldmath$\cal R$}$ is the Reynolds stress defined by Eq.~(\ref{eq:Re_str_def}), ${\bf{F}}$ the mean external force, $P$ the mean pressure function. If we substitute expression~(\ref{eq:Re_str_exp}) for $\mbox{\boldmath$\cal R$}$ into Eq.~(\ref{eq:mean_U_eq_rot}), we have
\begin{eqnarray}
	\frac{\partial {\bf{U}}}{\partial t}
	&=& {\bf{U}} \times {\bf{\Omega}}
	+ {\bf{J} \times {\bf{B}}}
	+ \nu_{\rm{K}} \nabla^2 \left( {{\bf{U}} - \frac{\gamma}{\beta} {\bf{B}}} \right)
	+ {\bf{F}}\nonumber\\
	& & - \nabla \left( {
		P + \frac{1}{2} {\bf{U}}^2 
		+ \left\langle {\frac{1}{2} {\bf{b}}'{}^2} \right\rangle
		+ \frac{2}{3} K_{\rm{R}}
	} \right).
	\label{eq:mean_U_eq_exp}
\end{eqnarray}
Then, the mean vorticity equation is given by
\begin{eqnarray}
	\frac{\partial {\bf{\Omega}}}{\partial t}
	&=& \nabla \times \left[ {
	{\bf{U}} \times {\bf{\Omega}}
	+ {\bf{J} \times {\bf{B}}} \rule{0.ex}{3.ex}
	} \right.\nonumber\\
	& & \left. {
	+ \nu_{\rm{K}} \nabla^2 \left( {{\bf{U}} - \frac{\gamma}{\beta} {\bf{B}}} \right)
	} \right]
	+ \nabla \times {\bf{F}}.
	\label{eq:Omega_eq_exp}
\end{eqnarray}
In addition to the Reynolds stress, which contains fluctuating velocity correlation and the fluctuating Lorentz force $\langle {{\bf{j}}' \times {\bf{b}}'} \rangle$, the turbulence effects enter the mean velocity equation also through the mean Lorentz force ${\bf{J}} \times {\bf{B}}$. The mean Ohm's law is written as
\begin{equation}
	{\bf{J}}
	= \sigma \left( {
	{\bf{E}}
	+ {\bf{U}} \times {\bf{B}} 
	+ {\bf{E}}_{\rm{M}}
	} \right).
	\label{eq:mean_Ohms_law}
\end{equation}
If we substitute expression (\ref{eq:E_M_exp}) for the turbulent electromotive force into Eq.~(\ref{eq:mean_Ohms_law}) and solve it with respect to ${\bf{J}}$, we have the mean electric-current density as
\begin{equation}
	{\bf{J}}
	= \frac{1}{\beta} \left( {
	{\bf{U}} \times {\bf{B}} 
	+ \alpha {\bf{B}}
	+ \gamma \mbox{\boldmath$\Omega$} 
	- \frac{\partial {\bf{A}}}{\partial t}
	} \right)
	\label{eq:mean_J_exp}
\end{equation}
(${\bf{A}}$: vector potential). Here, we have dropped the molecular magnetic diffusivity $\eta$ as compared with the turbulent one $\beta$ ($\eta \ll \beta$). Then, the mean Lorentz force is expressed as
\begin{equation}
	{\bf{J}} \times {\bf{B}}
	= \frac{1}{\beta} 
	\left( {{\bf{U}} \times {\bf{B}}} \right) \times {\bf{B}}
	+ \frac{\gamma}{\beta}
	\mbox{\boldmath$\Omega$} \times {\bf{B}}
	- \frac{1}{\beta} \frac{\partial {\bf{A}}}{\partial t} \times {\bf{B}}.
	\label{eq:mean_Lorentz_force_exp}
\end{equation}
Here, we should note that the $\gamma$- or cross-helicity-related term remains in the mean Lorentz force expression. This gives a strong contrast to the $\alpha$- or helicity-related term, which disappears due to the alignment of $\alpha {\bf{B}}$ with ${\bf{J}}$ in Eq.~(\ref{eq:mean_J_exp}). In other words, unlike the turbulent helicity, the turbulent cross helicity contributes to the mean velocity or vorticity equation through the mean Lorentz force.

	Substituting Eq.~(\ref{eq:mean_Lorentz_force_exp}) into the mean Lorentz force in Eq.~(\ref{eq:Omega_eq_exp}), we obtain the equation for the mean vorticity as
\begin{eqnarray}
	\frac{\partial {\bf{\Omega}}}{\partial t}
	&=& \nabla \times \left[ {
	\left( {
	{\bf{U}} - \frac{\gamma}{\beta} {\bf{B}}
	} \right)
	\times {\bf{\Omega}}
	+ \nu_{\rm{K}} \nabla^2 \left( {
	{\bf{U}} - \frac{\gamma}{\beta} {\bf{B}}
	} \right)
	} \right]
	\nonumber\\
	&+& \nabla \times \left[ {
	{\bf{F}}
	+ \frac{1}{\beta} 
	\left( {{\bf{U}} \times {\bf{B}}} \right) \times {\bf{B}}
	- \frac{1}{\beta} \frac{\partial {\bf{A}}}{\partial t} \times {\bf{B}}
	} \right].
	\label{eq:mean_Omega_eq_exp}
\end{eqnarray}

	This mean vorticity equation was originally derived in the context of system rotation effect.\cite{ito2005} As long as the expansion (\ref{eq:U_expand}) is adopted and the induced vorticity is considered, this equation can be also utilized for cases without a system rotation.



\end{document}